\documentclass[aps,pra,showpacs,twocolumn]{revtex4-1}
\usepackage{}
\usepackage{amsmath}
\usepackage{graphicx}
\usepackage{float}
\usepackage{amssymb}
\usepackage{graphicx}
\usepackage{epstopdf}
\usepackage{times,txfonts}
\usepackage{easybmat}
\usepackage{mathtools}
\newcommand{\ket}[1]{|#1\rangle}
\newcommand{\bra}[1]{\langle #1|}
\usepackage[bookmarks=false]{hyperref}
\hypersetup{colorlinks=true,citecolor=blue,linkcolor=blue,urlcolor=blue,pdfstartview=FitH,bookmarksopen=true}
\usepackage{color,soul}

\begin{document}
\title{Robust population transfer of spin states by geometric formalism}
\author{K. Z. Li}
\affiliation{Department of Physics, Shandong University, Jinan 250100, China}
\author{G. F. Xu}
\email{xgf@sdu.edu.cn}
\affiliation{Department of Physics, Shandong University, Jinan 250100, China}
\date{\today}
\begin{abstract}
Accurate population transfer of uncoupled or weakly coupled spin states is crucial for many quantum information processing tasks. In this paper, we propose a fast and robust scheme for population transfer which combines invariant-based inverse engineering and geometric formalism for robust quantum control. Our scheme is not constrained by the adiabatic condition and therefore can be implemented fast. It can also effectively suppress the dominant noise in spin systems, which together with the fast feature guarantees the accuracy of the population transfer. Moreover, the control parameters of the driving Hamiltonian in our scheme are easy to design because they correspond to the curvature and torsion of a three-dimensional visual space curve derived by using geometric formalism for robust quantum control. We test the efficiency of our scheme by numerically simulating the ground-state population transfer in $^{15}$N nitrogen vacancy centers and comparing our scheme with stimulated Raman transition, stimulated Raman adiabatic passage and conventional shortcuts to adiabaticity based schemes, three types of popularly used schemes for population transfer. The numerical results clearly show that our scheme is advantageous over these previous ones.
\end{abstract}
\maketitle

\section{Introduction}

As a fundamental module of quantum coherent control, accurate population transfer of spin states is the prerequisite for many quantum information processing tasks. To implement population transfer, people may first think of using Rabi oscillations. While Rabi oscillations are a convenient tool for population transfer, it can not handle the more challengeable situation where population transfer needs to be implemented between uncoupled or weakly coupled spin states. In such a situation, direct population transfer is forbidden and therefore Rabi oscillations are no longer used. To cope with this more challengeable situation, people resort to using an intermediate state to connect the two uncoupled or weakly coupled spin states, resulting in three-level system based population transfer schemes. Compared to the population transfer using Rabi oscillations, three-level system based population transfer is more difficult to keep accurate, and therefore particular methods need to be developed to make sure the quality of the transfer is satisfactory.

Two well-known population transfer schemes for uncoupled or weakly coupled states are stimulated Raman transition (SRT) \cite{SRT1,SRT2,SRT3} and stimulated Raman adiabatic passage (STIRAP) \cite{STI1,STI2,STI3,STI4,STI5}. Whilst these two schemes have been proven very efficient, there is still room left for improvement. For SRT, it is technically easy to realize and not constrained by the adiabatic condition. But it is sensitive to the frequency errors resulting from the fluctuation of the magnetic field \cite{bohm} that is ubiquitous and dominant in spin systems such as nitrogen vacancy centers in diamond \cite{wanghl1,wanghl2,cai1,bohm,tian} and semiconductor quantum dots \cite{LZD3,LZD4,LZD5,LZD6,LZD7}. On the other hand, STIRAP is insensitive to the frequency errors, which is an absolutely attracting feature, but it requires the quantum system to evolve adiabatically. It is known that adiabatic evolutions require long run time \cite{tong,cp2,optim0} and this makes STIRAP vulnerable to environment-induced decoherence \cite{STI4}. Recently, shortcuts to adiabaticity (STA) \cite{NSTA1,NSTA2}, which includes transitionless quantum driving, invariant-based inverse engineering and fast-forward approaches, has been used to speed up adiabatic population transfers \cite{YXX,bzhou1,Daems,chenxi20122,BZFL,FZB,song2016,dress2016,xia2015,xia2014,xiay2017,duyx2016,chenxi2016,chenxi2012,chenxi2010,inv1,optim1,DSEP1,DSEP2,DSEP3}
and design stimulated Raman exact passage \cite{ssp2,noptim2,liubj1,liubj0,XKS,xl22}. However, when using such STA based schemes in spin systems, the existence of the frequency errors still influences the performance of these schemes.

In this paper, we propose a robust scheme for population transfer between uncoupled or weakly coupled spin states. Our scheme combines invariant-based inverse engineering of STA and geometric formalism for robust quantum control. Geometric formalism for robust quantum control \cite{gh5,gh13,gh12,gh2,gh3,gh4,gh1} is used to suppress the frequency errors resulting from the fluctuation of the magnetic field and for simplicity will be sometimes referred to as geometric formalism in the following. Our scheme has two attracting features: fast implementation and robustness against the frequency errors. Considering fast implementation can reduce the influence of decoherence and the fluctuation of the magnetic field is the dominant noise in spin systems, our scheme has the potential to transfer the population of uncoupled or weakly coupled spin states accurately. Besides the above two features, our scheme is also friendly in experiment. The control parameters of our driving Hamiltonian can be designed by analyzing the curvature and torsion of a three-dimensional space curve that is derived using geometric formalism. We demonstrate the specific realization procedure of our scheme by numerically simulating the ground-state spin transfer in the $^{15}$N nitrogen vacancy center. We also compare our scheme with SRT, STIRAP and conventional STA based schemes, respectively, and the results show that our scheme is advantageous over these previous ones.

\section{theoretical framework}

\begin{figure}
  \includegraphics[scale=0.16]{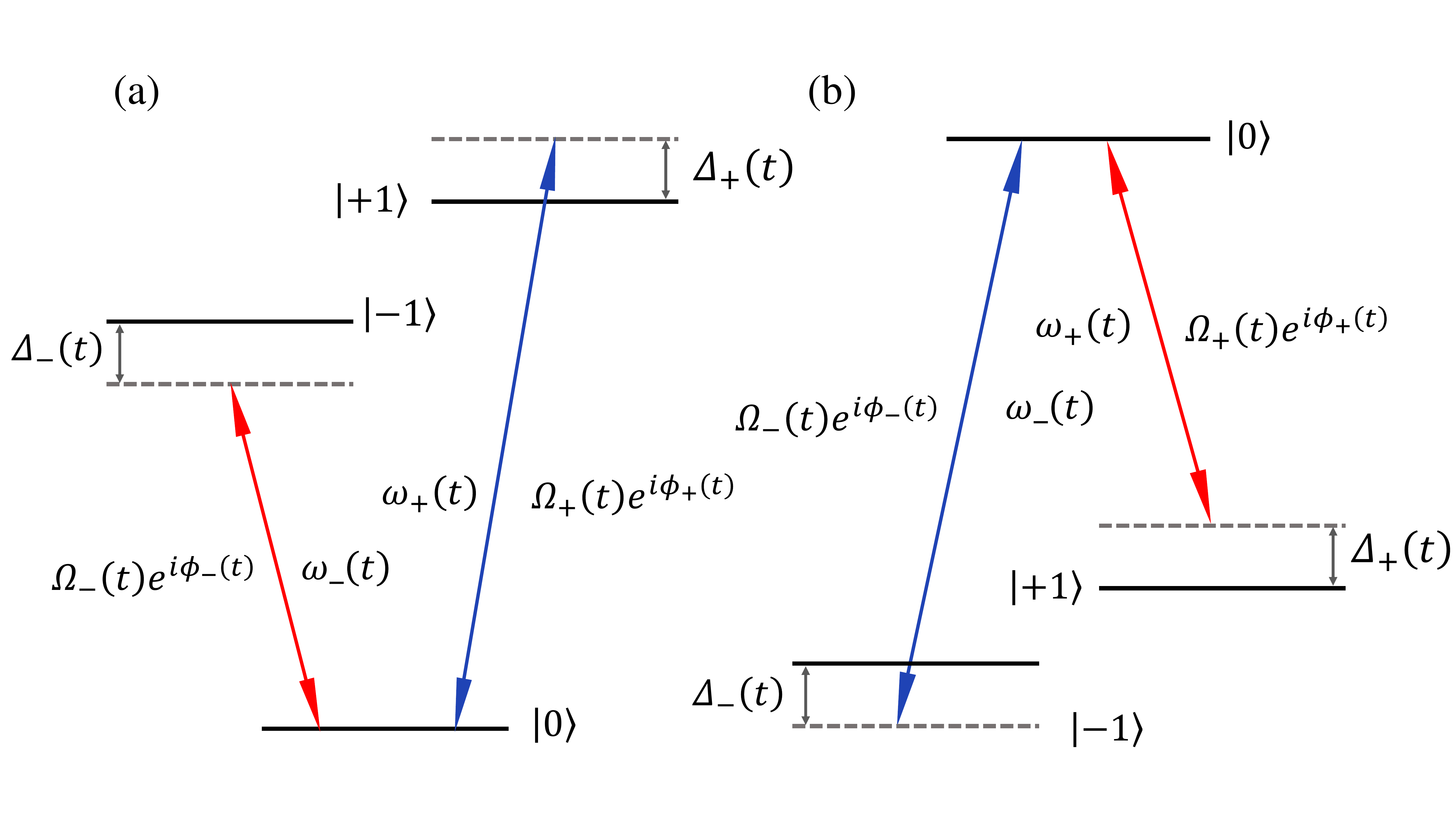}
  \caption{$V$-type and $\Lambda$-type three-level systems and the corresponding driving fields applied to them. $\Omega_{-}(t)$ and $\Omega_{+}(t)$ are the Rabi frequencies of the pump and Stokes fields, $\phi_{-}(t)$ and $\phi_{+}(t)$ are their phases, and $\omega_{-}(t)$ and $\omega_{+}(t)$ are their frequencies. $\Delta_{-}(t)$ and $\Delta_{+}(t)$ are the detunings from the resonances.}
   \label{Fig1}
\end{figure}
We illustrate our theoretical framework in this section. Consider a three-level spin system with states $\ket{-1}$, $\ket{+1}$ and $\ket{0}$. While the transitions $\ket{0}\leftrightarrow\ket{\pm1}$ are dipole coupled, the transition $\ket{-1}\leftrightarrow\ket{+1}$ is dipole forbidden. So, the system we consider can have either a $V$ structure depicted by Fig. \ref{Fig1} (a) or a $\Lambda$ structure depicted by Fig. \ref{Fig1} (b). The population transfer between spin states $\ket{-1}$ and $\ket{+1}$ is required to implement, but since the transition $\ket{-1}\leftrightarrow\ket{+1}$ is dipole forbidden, the state $\ket{0}$ is used as an intermediate state. Our theoretical framework is suitable for both the $V$-type and $\Lambda$-type spin systems, and without loss of generality, we use the $V$-type spin system to do the illustration. Our aim is to realize accurate population transfer from $\ket{-1}$ to $\ket{+1}$ in the presence of noise. To achieve the aim, we consider the following form of Hamiltonian
\begin{align}\label{eq2}
H(t)=\left(
       \begin{array}{ccc}
         \Delta(t) & \frac{1}{\sqrt{2}}\Omega(t)e^{-i\phi(t)} & 0\\
         \frac{1}{\sqrt{2}}\Omega(t)e^{i\phi(t)} & 0 & \frac{1}{\sqrt{2}}\Omega(t)e^{-i\phi(t)}\\
         0 & \frac{1}{\sqrt{2}}\Omega(t)e^{i\phi(t)} & -\Delta(t)\\
       \end{array}
     \right).
\end{align}
The Hamiltonian $H(t)$ is written in the basis $\{\ket{-1}, \ket{0}, \ket{+1}\}$, and it can be realized by applying the driving fields shown in Fig. \ref{Fig1} (a) to the system. In Fig. \ref{Fig1} (a), $\Omega_{-}(t)$ and $\Omega_{+}(t)$ are the Rabi frequencies of the pump and Stokes fields, respectively, and we assume that they have the same envelope $\Omega_{-}(t)=\Omega_{+}(t)=\sqrt{2} \Omega(t)$, where the constant $\sqrt{2}$ is just for the convenience of subsequent calculations. $\phi_{-}(t)$ and $\phi_{+}(t)$ are the phases of these two driving fields, and they have the same value all the time, that is, $\phi_{-}(t)=-\phi_{+}(t)=\phi(t)$. The detunings of these two driving fields are denoted by $\Delta_{-}(t)=(E_{-1}-E_{0})-\omega_{-}(t)$ and $\Delta_{+}(t)=(E_{+1}-E_{0})-\omega_{+}(t)$, respectively, and we also assume they have the same value $\Delta_{-}(t)=-\Delta_{+}(t)=\Delta(t)$, where $\omega_{-}(t)$ and $\omega_{+}(t)$ are the corresponding frequencies of the pump and Stokes fields, and $E_{-1}$, $E_{0}$ and $E_{+1}$ are the bare-basis state energies. It's worth noting that our scheme also applies to effective three-level systems such as two interacting spins \cite{inv1,DSEP1}.

Note that with $\Omega(t)$, $\Delta(t)$ and $\phi(t)$ being different functions with respect to time $t$, the Hamiltonian $H(t)$ in Eq.~(\ref{eq2}) will be different. What we will do is to give an approach to set the functions $\Omega(t)$, $\Delta(t)$ and $\phi(t)$, making the population transfer from $\ket{-1}$ to $\ket{+1}$ accurate even under the influence of noise. The proposed approach combines invariant-based inverse engineering and geometric formalism for robust quantum control. Specifically, the inverse engineering in our approach is inspired by geometric formalism. The procedure of our approach is as follows. We first design the evolution operator $U(t,0)$ of the quantum system with the help of dynamical invariants. Then we analyse the influence of noise on the evolution operator $U(t,0)$ with Dyson series. During this process, geometric formalism is introduced and it turns the population transfer problem into a space curve design problem. The information about how to set the control parameters $\Omega(t)$, $\Delta(t)$ and $\phi(t)$ can be obtained from inversely calculating the curvature and torsion of the space curve. We in the following illustrate our approach in detail.

As illustrated above, we first design the evolution operator $U(t,0)$ induced by the Hamiltonian $H(t)$. Considering directly solving the time-dependent Schr\"{o}dinger equation $i\partial \ket{\psi(t)}/\partial t =H(t)\ket{\psi(t)}$ is hard, we will use the dynamical invariant $I(t)$ related to $H(t)$ to parameterize the evolution operator $U(t,0)$, i.e., to express the evolution operator $U(t,0)$ with some other parameters instead of $\Omega(t)$, $\Delta(t)$ and $\phi(t)$ in the Hamiltonian $H(t)$. Note that although we do not give the expression of $U(t,0)$ in terms of $\Omega(t)$, $\Delta(t)$ and $\phi(t)$, it is sufficient for our subsequent discussion.

To parameterize the evolution operator $U(t,0)$, we rewrite the Hamiltonian $H(t)$ by expanding it with spin-1 angular momentum operators,
\begin{align}
H(t)=\Omega(t)\cos\phi(t)K_{x}+\Omega(t)\sin\phi(t)K_{y}+\Delta(t)K_{z},
     \label{eq3}
\end{align}
where the three spin-1 angular momentum operators in the basis $\{\ket{-1}, \ket{0}, \ket{+1}\}$ read
\begin{align}
K_{x}=\frac{1}{\sqrt{2}}\left(
       \begin{array}{ccc}
         0 & 1 & 0\\
         1 & 0 & 1\\
         0 & 1 & 0\\
       \end{array}
     \right),
K_{y}=\frac{1}{\sqrt{2}}\left(
       \begin{array}{ccc}
         0 & -i & 0\\
         i & 0 & -i\\
         0 & i & 0\\
       \end{array}
     \right),
K_{z}=\left(
       \begin{array}{ccc}
         1 & 0 & 0\\
         0 & 0 & 0\\
         0 & 0 & -1\\
       \end{array}
     \right).
     \label{eq4}
\end{align}
One can verify that the operators $K_{x}$, $K_{y}$ and $K_{z}$ form a closed algebra and this algebra is isomorphic to the Lie algebra of SU(2), i.e., their commutation relations satisfy
\begin{align}
[K_{x},K_{y}]=iK_{z},[K_{y},K_{z}]=iK_{x},[K_{z},K_{x}]=iK_{y}.
\label{eq5}
\end{align}
From the above, one can see that the Hamiltonian $H(t)$ in Eq.~(\ref{eq2}) possesses SU(2) dynamical symmetry. Because of this feature, the relevant dynamical invariant $I(t)$, such that $dI(t)/dt\equiv\partial I(t)/\partial t-i[I(t),H(t)]=0$, can be constructed as \cite{inv1,inv2}
\begin{align}
I(t)=&\Omega_{0}[\cos\beta(t)\sin\theta(t) K_{x}+\sin\beta(t)\sin\theta(t) K_{y} \nonumber \\
&+\cos\theta(t) K_{z}],
     \label{eq51}
\end{align}
where $\Omega_{0}$ is an arbitrary constant with units of frequency, guaranteeing $I(t)$ having the same energy dimension as $H(t)$, and $\beta(t)$ and $\theta(t)$ are time-dependent parameters related to $\Omega(t)$, $\Delta(t)$ and $\phi(t)$. By solving the equation $I(t)\ket{\varphi_{n}(t)}=\lambda_{n}\ket{\varphi_{n}(t)}$, one can readily get the eigenvalues and eigenstates of $I(t)$. The eigenvalues are $\lambda_{1}=\Omega_{0}$, $\lambda_{2}=0$, $\lambda_{3}=-\Omega_{0}$, and the corresponding eigenstates are
\begin{align}\label{eq52}
\ket{\varphi_{1}(t)}&=\left(
       \begin{array}{c}
         \cos^{2}\frac{\theta(t)}{2}e^{-i\beta(t)}\\
         \frac{1}{\sqrt{2}}\sin\theta(t)\\
         \sin^{2}\frac{\theta(t)}{2}e^{i\beta(t)}\\
       \end{array}
     \right),\notag\\
\ket{\varphi_{2}(t)}&=\left(
       \begin{array}{c}
         -\frac{1}{\sqrt{2}}\sin\theta(t) e^{-i\beta(t)}\\
         \cos\theta(t)\\
         \frac{1}{\sqrt{2}}\sin\theta(t) e^{i\beta(t)}\\
       \end{array}
     \right),\\
\ket{\varphi_{3}(t)}&=\left(
       \begin{array}{c}
         \sin^{2}\frac{\theta(t)}{2}e^{-i\beta(t)}\\
         -\frac{1}{\sqrt{2}}\sin\theta(t)\\
          \cos^{2}\frac{\theta(t)}{2}e^{i\beta(t)}\\
       \end{array}
     \right).\notag
\end{align}
According to Lewis-Riesenfeld theory, instead of directly solving the time-dependent
Schr\"{o}dinger equation $i\partial \ket{\psi(t)}/\partial t =H(t)\ket{\psi(t)}$, which is hard, the solution to it can be expanded by the orthonormal dynamical modes $e^{i\alpha_{n}(t)}\ket{\varphi_{n}(t)}$\cite{inv2}, that is,
\begin{align}
\ket{\psi(t)}=\sum_{n=1}^3C_{n}e^{i\alpha_{n}(t)}\ket{\varphi_{n}(t)}.
     \label{eq53}
\end{align}
In the above, $C_{n}$ are time-independent amplitudes, $\ket{\varphi_{n}(t)}$ are the eigenstates of the invariant $I(t)$, and the phases
\begin{align}
\alpha_{n}(t)=\int^{t}_{0}\bra{\varphi_{n}(t^{\prime})}i\frac{\partial}{\partial t^{\prime}}-H(t^{\prime})\ket{\varphi_{n}(t^{\prime})}dt^{\prime}.
     \label{eq54}
\end{align}
Moreover, by calculation one can get that $\alpha_{2}(t)=0$ and $\alpha_{1}(t)$ is always equal to $-\alpha_{3}(t)$, i.e., $\alpha_{1}(t)=-\alpha_{3}(t)=\alpha(t)$ with $\alpha(t)$ being the common value. From Eq.~(\ref{eq53}), one can see that the evolution driven by $H(t)$ can be divided into three orthonormal dynamical modes $e^{i\alpha_{n}(t)}\ket{\varphi_{n}(t)}$ with $n=1,2,3$. Considering analysing one single dynamical mode is easier than analysing the interference of these modes and our aim is to realize the population transfer from $\ket{-1}$ to $\ket{+1}$, we here set $\ket{\varphi_{1}(0)}=(1,0,0)^{\mathcal {T}}$, $\ket{\varphi_{2}(0)}=(0,1,0)^{\mathcal {T}}$ and $\ket{\varphi_{3}(0)}=(0,0,1)^{\mathcal {T}}$, planning to let the transition from $\ket{-1}$ to $\ket{+1}$ evolve along the first dynamical mode. Note that $\mathcal {T}$ in the above means the transpose of matrixes and the states $\ket{\varphi_{n}(0)}$ are written in the basis $\{\ket{-1}, \ket{0}, \ket{+1}\}$. With the calculated values of $\alpha_{n}(t)$ and the setting of $\ket{\varphi_{n}(0)}$, the evolution operator $U(t,0)$ can be written as
\begin{align}\label{eq6}
\begin{split}
U(t,0)&=\sum_{n=1}^3e^{i\alpha_{n}(t)}\ket{\varphi_{n}(t)}\bra{\varphi_{n}(0)}\\
&=\left(\begin{array}{ccc}
         \cos^{2}\frac{\theta}{2}e^{-i(\beta-\alpha)}& -\frac{1}{\sqrt{2}}\sin\theta e^{-i\beta} & \sin^{2}\frac{\theta}{2}e^{-i(\beta+\alpha)}\\
         \frac{1}{\sqrt{2}}\sin\theta e^{i\alpha} & \cos\theta & -\frac{1}{\sqrt{2}}\sin\theta e^{-i\alpha}\\
         \sin^{2}\frac{\theta}{2}e^{i(\beta+\alpha)} & \frac{1}{\sqrt{2}}\sin\theta e^{i\beta} & \cos^{2}\frac{\theta}{2}e^{i(\beta-\alpha)}\\
       \end{array}
     \right),
\end{split}
\end{align}
where $\alpha$, $\beta$ and $\theta$ have been used to represent $\alpha(t)$, $\beta(t)$ and $\theta(t)$ for conciseness.

Until now, we have parameterized the evolution operator $U(t,0)$ by expressing it with $\alpha(t)$, $\beta(t)$ and $\theta(t)$. We next will analyse the influence of noise on the population transfer with the help of Dyson series, and based on the analysis, give an approach to realize accurate population transfer. Specifically, we expand the practical final state to the second order with the help of Dyson series and $U(t,0)$ in Eq.~(\ref{eq6}), and define a space curve which can describe the evolution of the system.

For the three-level spin system, the dominant noise is the fluctuation of the magnetic field, which results from the influence of environments and the imperfection of magnetic control. Generally, the fluctuation of the magnetic field is much slower than the typical operate time, allowing one to use the quasistatic noise model to describe it. Moreover, due to the linear dependence of the energies $\ket{\pm1}$ on the magnetic field, the influence of the fluctuation of the magnetic field can be further seen as frequency errors. After accounting into the dominant noise which can be treated as frequency errors, the Hamiltonian of the three-level spin system turns into
\begin{align}
H^{\prime}(t)=H(t)+\delta K_{z},
     \label{eq 7}
\end{align}
where $\delta$ represents the strength of the frequency errors, and it is sufficiently small compared to those of the driving fields. In this case, the term $\delta K_{z}$ can be seen as a perturbation to the Hamiltonian $H(t)$. By using Dyson series, we expand the practical final state $\ket{\psi^{\prime}(T)}$ to the second-order,
\begin{align} \label{eq 71}
\ket{\psi^{\prime}(T)}&=\ket{\psi(T)}-i\delta\int^{T}_{0}dtU(T, t)K_{z}\ket{\psi(t)}\\
&-\delta^{2}\int^{T}_{0}dt\int^{t}_{0}dt^{\prime}U(T, t)K_{z}U(t, t^{\prime})K_{z}\ket{\psi(t^{\prime})}+\cdots,\notag
\end{align}
where $\ket{\psi(t)}$ is the unperturbed state and $U(s,t)=\sum_{n}\ket{\psi_{n}(s)}\bra{\psi_{n}(t)}$ is the unperturbed evolution operator, with $\ket{\psi_{n}(t)}=e^{i\alpha_{n}(t)}\ket{\varphi_{n}(t)}$ representing the orthonormal dynamical modes. Recall that we plan to let the transition from states $\ket{-1}$ to $\ket{+1}$ evolve along the first dynamical mode, which means the unperturbed state $\ket{\psi(t)}$ would be $\ket{\psi_{1}(t)}$. With the unperturbed final state described by $\ket{\psi_{1}(T)}$ and the practical final state $\ket{\psi^{\prime}(T)}$ described by Eq.~(\ref{eq 71}), the quality of the population transfer can be assessed by the fidelity $F=|\langle\psi_{1}(T)|\psi^{\prime}(T)\rangle|^{2}$ and it reads
\begin{align} \label{eq 72}
F\approx1-\delta^{2}\sum_{n\neq 1}\big|\int_{0}^{T}dt
\bra{\psi_{1}(t)}K_{z}\ket{\psi_{n}(t)}\big|^{2},
\end{align}
where the second term is the noise term, reflecting the influence of the frequency errors on the population transfer. From the above equation, one can see that if the noise term $\sum_{n\neq 1}\big|\int_{0}^{T}dt\bra{\psi_{1}(t)}K_{z}\ket{\psi_{n}(t)}\big|^{2}$ can be suppressed, the fidelity will approach to one and accurate population transfer in the presence of noise can be realized. For the convenience of subsequent discussions, we define an operator $m(t)$ as
\begin{align} \label{eq 75}
m(t)=\int^{t}_{0}U^{\dag}(t^{\prime},0)K_{z}U(t^{\prime},0)dt^{\prime}.
\end{align}
One can verify that if $m(T)=0$, the noise term $\sum_{n\neq 1}\big|\int_{0}^{T}dt\bra{\psi_{1}(t)}K_{z}\ket{\psi_{n}(t)}\big|^{2}$  will be suppressed. Substituting Eqs.~(\ref{eq4}) and (\ref{eq6}) into Eq.~(\ref{eq 75}), one finds that the operator $m(t)$ can also be expanded with spin-1 angular momentum operators, that is
\begin{align}
m(t)=x(t)K_{x}+y(t)K_{y}+z(t)K_{z}=\mathbf{r}(t)\cdot\hat{K}.
\label{eq10}
\end{align}
In the above, $x(t)=\frac{1}{2}\text{Tr}(K_xm(t))$, $y(t)=\frac{1}{2}\text{Tr}(K_ym(t))$, $z(t)=\frac{1}{2}\text{Tr}(K_zm(t))$ are the expansion coefficients, $K_{x}$, $K_{y}$, $K_{z}$ are spin-1 angular momentum operators described by Eq. (\ref{eq4}). Correspondingly, $\mathbf{r}(t)=(x(t), y(t), z(t))$ is the coefficient vector and $\hat{K}=(K_{x},K_{y},K_{z})$ is the vector composed of spin-1 angular momentum operators. The above equation tells us that the operator $m(t)$ can be totally described by the three-dimensional space curve defined by
\begin{align}\label{eq101}
\mathbf{r}(t)=x(t)\hat{e_x}+ y(t)\hat{e_y}+ z(t)\hat{e_z},
\end{align}
because $K_x$, $K_y$ and $K_z$ are fixed matrixes, where $\hat{e_x}$, $\hat{e_y}$ and $\hat{e_z}$ are orthonormal vectors in the three-dimensional Euclidean space.

The space curve $\mathbf{r}(t)$ defined above induces geometric formalism for robust population transfer and by using geometric formalism, we obtain the main result of our paper. To realize robust population transfer, one needs to satisfy two conditions. Condition (i): the noise term in Eq.~(\ref{eq 72}) can be suppressed, i.e., $m(T)=0$. Condition (ii): the desired population transfer is realized in the ideal case, i.e., the transfer from $\ket{-1}$ to $\ket{+1}$ can be driven by the Hamiltonian $H(t)$. It turns out that these two conditions both can turn into conditions on the space curve $\mathbf{r}(t)$. More importantly, by further calculating the curvature and torsion of the space curve $\mathbf{r}(t)$, one can get $\Omega(t)$, $\Delta(t)$ and $\phi(t)$.

We first show that condition (i) can turn into a condition on the space curve $\mathbf{r}(t)$. Since $m(0)=0$, the space curve starts at the origin $\mathbf{r}(0)=0$. According to Eq.~(\ref{eq 75}), the condition $m(T)=0$ turns into $\mathbf{r}(T)=0$. This condition tell us that if the space curve $\mathbf{r}(t)$ is closed, the noise term $\sum_{n\neq 1}\big|\int_{0}^{T}dt\bra{\psi_{1}(t)}K_{z}\ket{\psi_{n}(t)}\big|^{2}$ can be suppressed. We next discuss condition (ii). To this end, we apply the parameterized evolution operator $U(t,0)$ in Eq.~(\ref{eq6}) to the initial state $\ket{-1}$. As is desired, the system evolves along the first dynamical mode $\ket{\psi_{1}(t)}$, that is,
\begin{align}
\ket{\psi(t)}=\ket{\psi_{1}(t)}=e^{i\alpha(t)}\left(
       \begin{array}{ccc}
         \cos^{2}\frac{\theta(t)}{2}e^{-i\beta(t)}\\
         \frac{1}{\sqrt{2}}\sin\theta(t) \\
         \sin^{2}\frac{\theta(t)}{2}e^{i\beta(t)} \\
       \end{array}
     \right).
\label{eq19}
\end{align}
To fulfill the population transfer, the final state $\ket{\psi(T)}$ should be $\ket{+1}$. This gives us the condition
\begin{align}
\theta(0)=0,~~~~~\theta(T)=\pi.
\label{eq20}
\end{align}
The above condition does not put special conditions on $\mathbf{r}(t)$, but it constrains $\dot{\mathbf{r}}(t)$, the derivative of $\mathbf{r}(t)$ with respect to time $t$. To see this, one can use Eqs. (\ref{eq  75}-\ref{eq101}) to calculate $\dot{\mathbf{r}}(t)$ and it reads
\begin{align}
\dot{\mathbf{r}}(t)=-\sin\theta(t)\cos\alpha(t)\hat{e_x}-\sin\theta(t)\sin\alpha(t)\hat{e_y}
+\cos\theta(t)\hat{e_z}.
\label{eq11}
\end{align}
The length of $\dot{\mathbf{r}}(t)$ is unit, implying that $\dot{\mathbf{r}}(t)$ is the tangent vector of the curve and $\mathbf{r}(t)$ is the parametrization of the curve by arc length. Substituting Eq.~(\ref{eq20}) into Eq. (\ref{eq11}), one can see that the condition in Eq.~(\ref{eq20}) turns into
\begin{align}
\dot{\mathbf{r}}(0)=(0,0,1),~~~\dot{\mathbf{r}}(T)=(0,0,-1),
\label{eq21}
\end{align}
which means that the tangent vectors of the space curve $\mathbf{r}(t)$ at $t=0$ and $t=T$ are fixed.

In the previous paragraph, we have shown conditions (i) and (ii) can turn into conditions on the space curve $\mathbf{r}(t)$, that is, the space curve $\mathbf{r}(t)$ must be closed and the tangent vectors of the space curve $\mathbf{r}(t)$ at $t=0$ and $t=T$ should be $\dot{\mathbf{r}}(0)=(0,0,1)$ and $\dot{\mathbf{r}}(T)=(0,0,-1)$. In the following, we will show that by calculating the curvature and torsion of the space curve, one can get $\Omega(t)$, $\Delta(t)$ and $\phi(t)$.

Because Eq.~(\ref{eq 75}) contains the evolution operator $U(t,0)$, $m(t)$ contains all the information to describe the evolution. This point makes it possible to obtain $\Omega(t)$, $\Delta(t)$ and $\phi(t)$ by calculating the curvature and torsion of the space curve $\mathbf{r}(t)$. By calculation, the derivatives of $m(t)$ are
\begin{align}
\dot{m}(t)=\dot{\mathbf{r}}(t)\cdot\hat{K}=U^{\dag}(t,0)K_{z}U(t,0),
\label{eq12}
\end{align}
\begin{align}
\ddot{m}(t)=\ddot{\mathbf{r}}(t)\cdot\hat{K}=iU^{\dag}(t,0)[H(t),K_{z}]U(t,0),
\label{eq13}
\end{align}
\begin{align}\label{eq14}
\begin{split}
\dddot{m}(t)=\dddot{\mathbf{r}}(t)\cdot\hat{K}=-&U^{\dag}(t,0)H(t)[H(t),K_{z}]U(t,0)\\
+&iU^{\dag}(t,0)[\dot{H}(t),K_{z}]U(t,0)\\
+& U^{\dag}(t,0)[H(t),K_{z}]H(t)U(t,0).
\end{split}
\end{align}
Substituting Eq. (\ref{eq2}) into Eq. (\ref{eq13}), one can get that $\Omega(t)$ is equal to the curvature $\kappa(t)=\|\ddot{\mathbf{r}}(t)\|$ of the space curve $\mathbf{r}(t)$, i.e.,
\begin{align}\label{eq15}
\Omega(t)=\|[H(t),K_{z}]\|_{F}=\|\ddot{\mathbf{r}}(t)\|.
\end{align}
In the above, the scaled Frobenius norm of matrixes is defined as $\|m\|_{F}=\sqrt{\sum^{n}_{i,j}|m_{ij}|^{2}}/\sqrt{2}$, which is invariant under unitary equivalence transformations of $m$.
Next, using the expressions of $\dot{m}(t)$, $\ddot{m}(t)$ and $\dddot{m}(t)$ in Eqs. (\ref{eq12}-\ref{eq14}), one can obtain
\begin{align}\label{eq16}
\dot{\phi}(t)-\Delta(t)=-i\frac{\textmd{Tr}\{\dot{m}(t)\ddot{m}(t)\dddot{m}(t)\}}{\|[\dot{m}(t),\ddot{m}(t)]\|^{2}_{F}}.
\end{align}
The above equation can be rewritten by using the identities of spin-1 angular momentum operators:
\begin{align}\label{eq17}
\begin{split}
\dot{\phi}(t)-\Delta(t)=&-i\frac{\textmd{Tr}\{\dot{m}(t)\ddot{m}(t)\dddot{m}(t)\}}{\|[\dot{m}(t),\ddot{m}(t)]\|^{2}_{F}}\\
=&-i\frac{\frac{1}{2}\textmd{Tr}\{[\dot{m}(t),\ddot{m}(t)]\dddot{m}(t)\}}{\|[\dot{m}(t),\ddot{m}(t)]\|^{2}_{F}}\\
=&-i\frac{\frac{1}{2}\textmd{Tr}\{[i(\dot{\mathbf{r}}(t)\times\ddot{\mathbf{r}}(t))\cdot\hat{K}](\dddot{\mathbf{r}}(t)\cdot\hat{K})\}}{\|i(\dot{\mathbf{r}}(t)\times\ddot{\mathbf{r}}(t))\cdot\hat{K}\|^{2}_{F}} \\
=&\frac{(\dot{\mathbf{r}}(t)\times\ddot{\mathbf{r}}(t))\cdot\dddot{\mathbf{r}}(t)}{\|\dot{\mathbf{r}}(t)\times\ddot{\mathbf{r}}(t)\|^{2}} \end{split}
\end{align}
which is just the torsion $\tau(t)$ of the space curve.

\section{DISCUSSION OF ROBUSTNESS}

While the general idea of our paper is given in the above section, we will give a concrete example to demonstrate the feasibility of the general idea in this section. Specifically, we will demonstrate the specific realization procedure of our geometric formalism based robust population transfer scheme with the ground-state population transfer in the $^{15}$N nitrogen vacancy center. We also numerically simulate the performance of our scheme in practical scenarios and compare it with those of SRT, STIRAP and conventional STA based schemes.
\begin{figure}[htb]
  \includegraphics[scale=0.18]{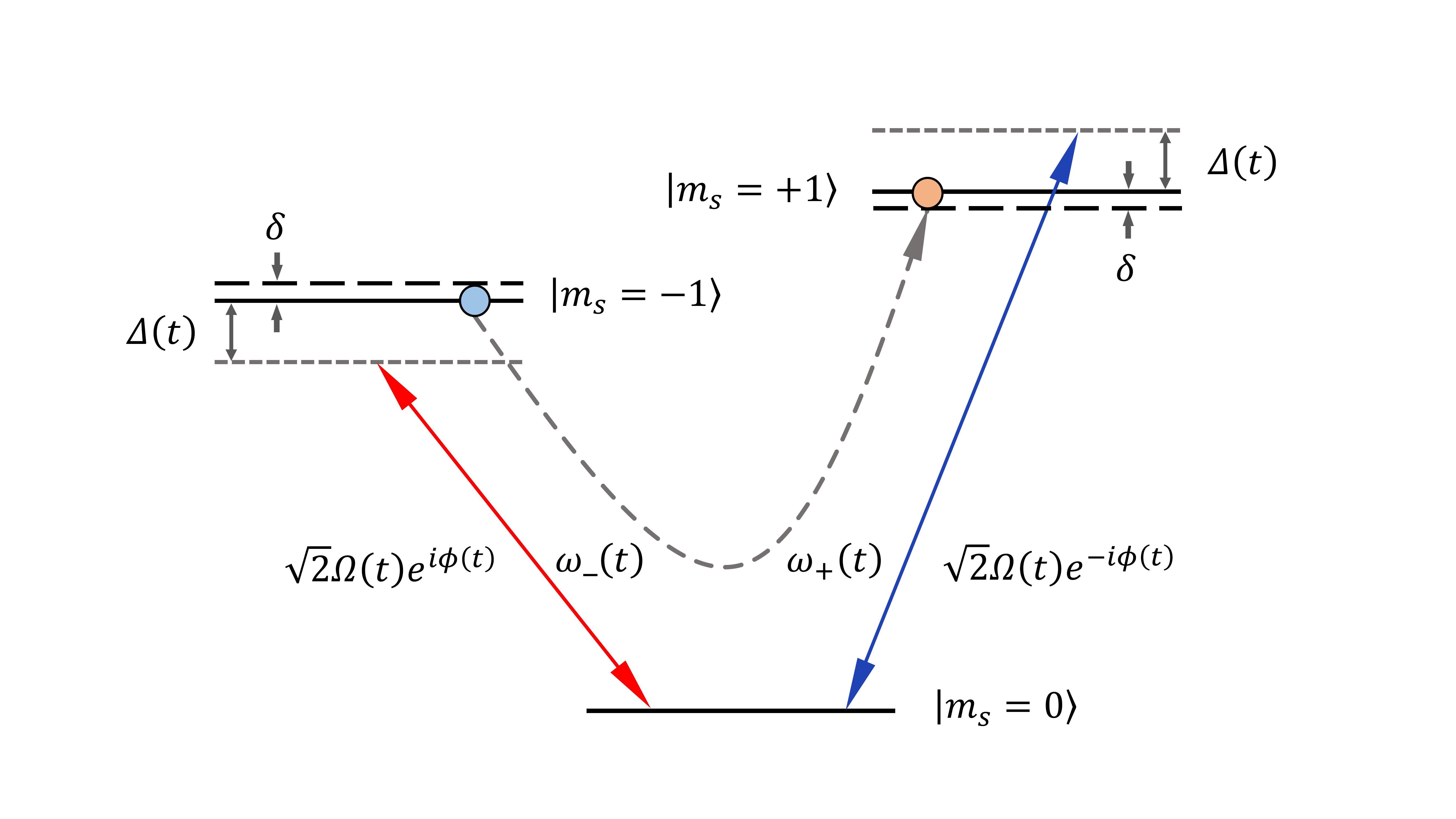}
  \caption{The energy diagram of the nitrogen vacancy center. Our aim is to drive the transition between states $\ket{m_{s}=-1}\leftrightarrow\ket{m_{s}=+1}$ with $\ket{m_{s}=0}$ as an intermediate state. $\delta$ describes the frequency errors resulting from the fluctuation of the magnetic field.}
   \label{Fig2}
\end{figure}
\begin{figure}[htb]
  \includegraphics[scale=0.5]{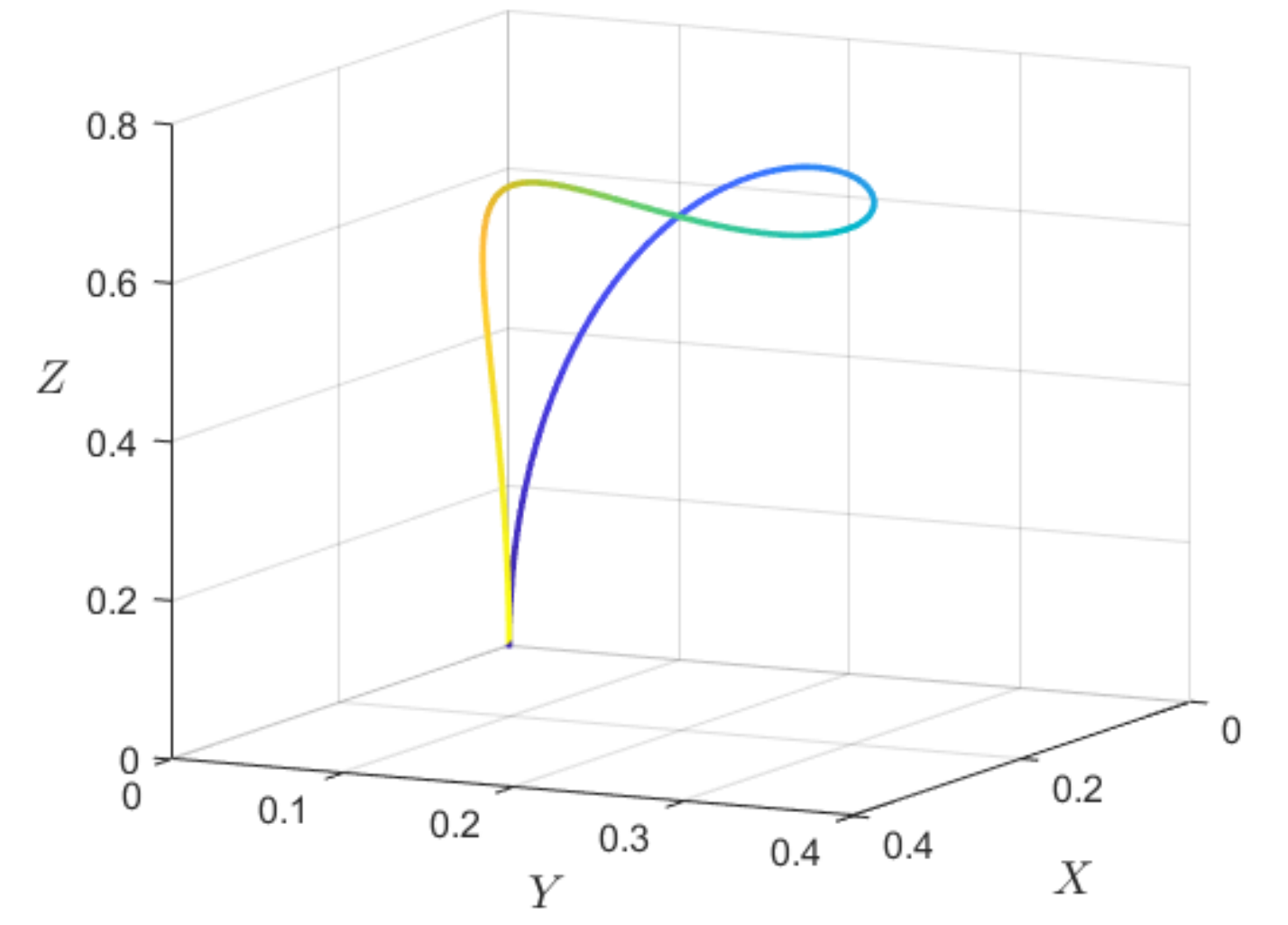}
  \caption{The curve $\textbf{r}(d)$ in the three-dimensional Euclidean space. The color changes from light to dark with the parameter increasing.}
   \label{Fig3}
\end{figure}

\begin{figure*}[htp]
  \includegraphics[scale=0.4]{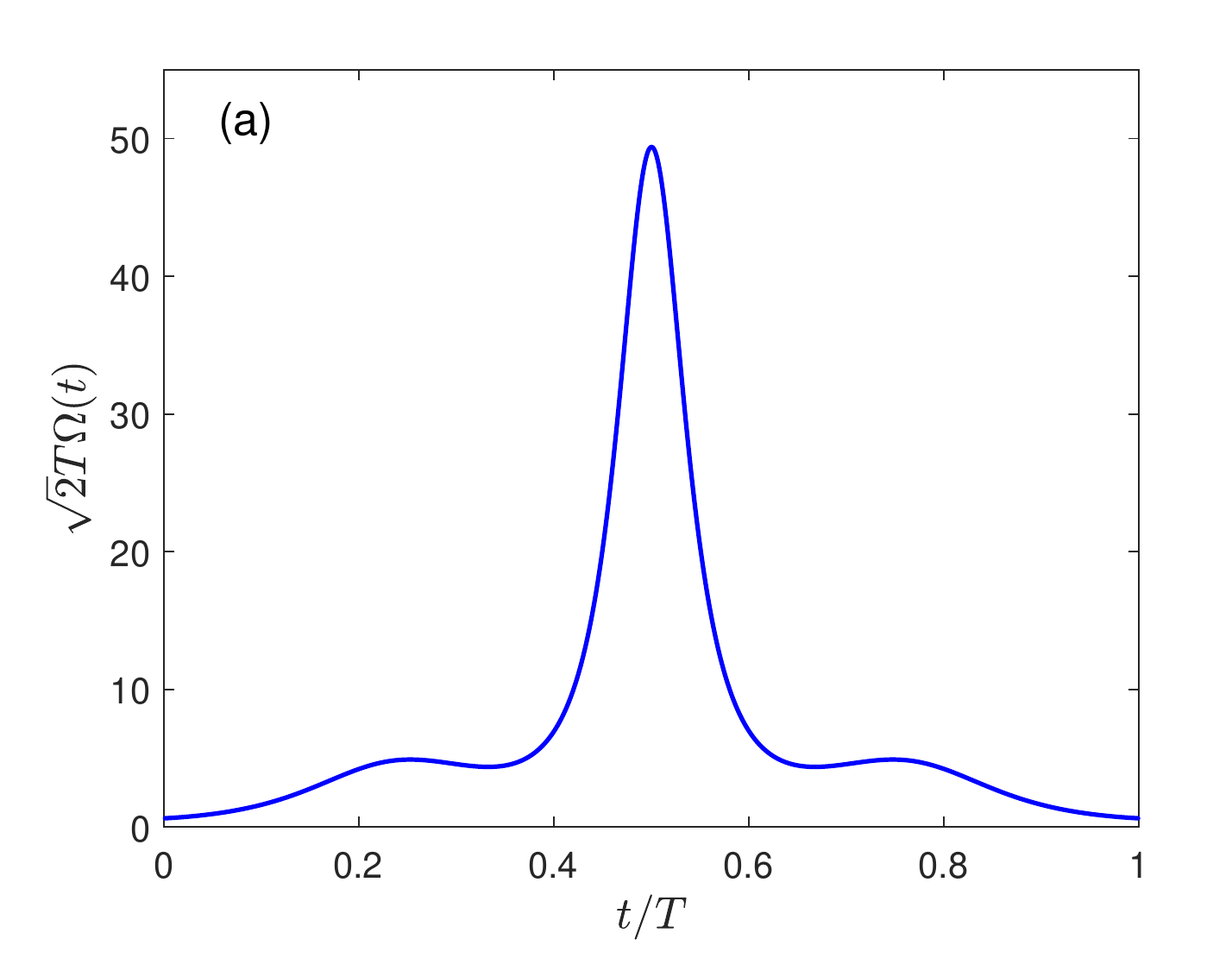}
  \includegraphics[scale=0.4]{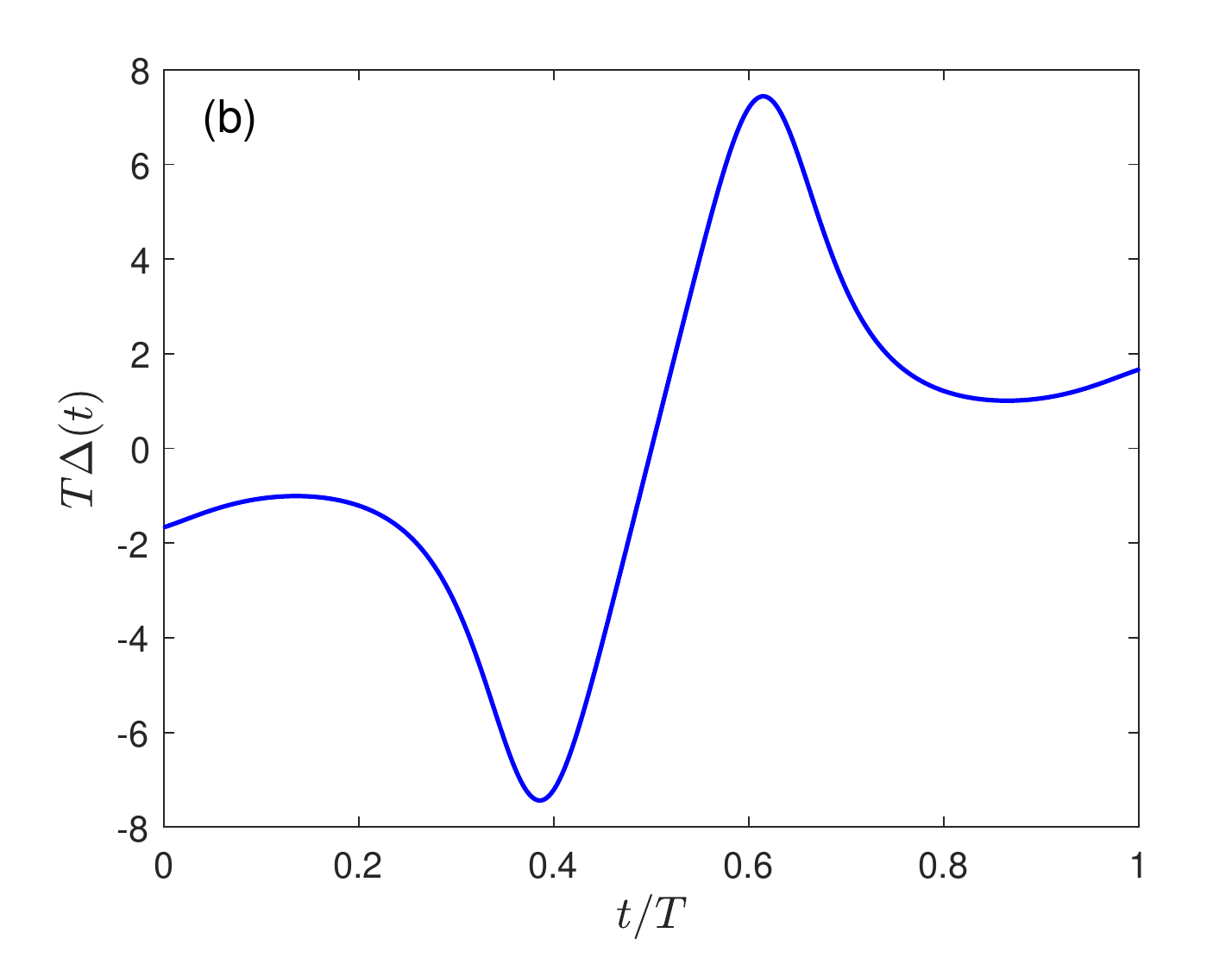}
  \includegraphics[scale=0.4]{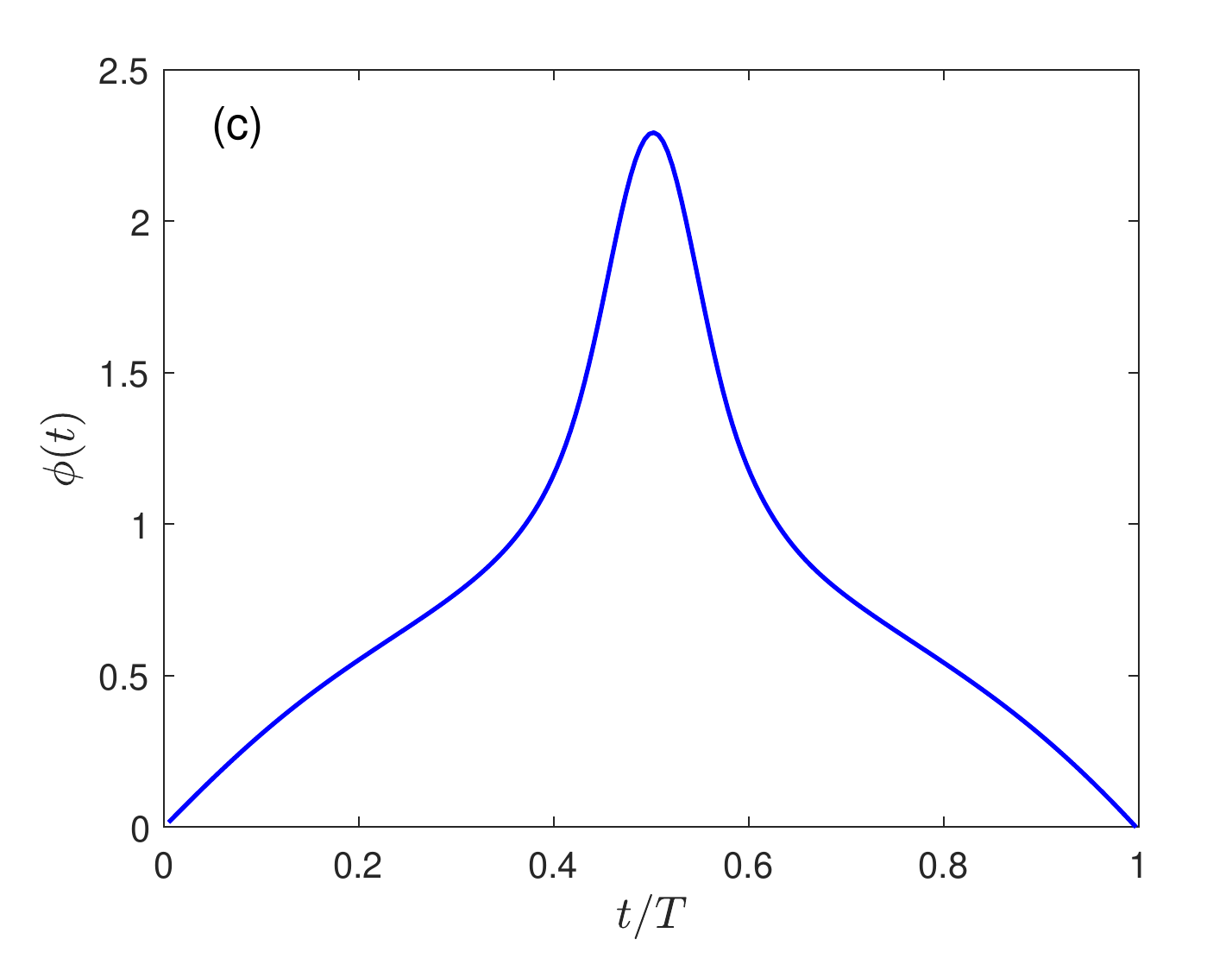}
  \caption{The control parameters of the driving Hamiltonian in our scheme. (a) The Rabi frequencies of the pump and Stokes fields $\Omega_{-}(t)=\Omega_{+}(t)=\sqrt{2} \Omega(t)$. (b) The detunings of the pump and Stokes fields $\Delta_{-}(t)=-\Delta_{+}(t)=\Delta(t)$. (c) The phases of the pump and Stokes fields $\phi_{-}(t)=-\phi_{+}(t)=\phi(t)$. The unit time $T$ is set as 2.116 $\mu$s corresponding to the arc length of the curve $\textbf{r}(t)$.}
   \label{Fig4}
\end{figure*}

Consider a $^{15}$N nitrogen vacancy center in the high-purity type IIa diamond whose host $^{15}\text{N}$ nuclear spin is polarized \cite{nuclpolar}. This system has a spin-triplet ground state $\ket{m_{s}=0}$ and $\ket{m_{s}=\pm1}$. The degeneracy between $\ket{m_{s}=\pm1}$ can be lifted by applying an external magnetic field $B_{z}$ along the symmetry axis of the nitrogen vacancy center, so the ground state of the nitrogen vacancy center can be described by a $V$-type system, as shown in Fig.~\ref{Fig2}, where the transitions $\ket{m_{s}=0}\leftrightarrow\ket{m_{s}=\pm1}$ are dipole coupled and the transition $\ket{m_{s}=-1}\leftrightarrow\ket{m_{s}=+1}$ is dipole forbidden. Our aim is to realize the population transfer from $\ket{m_{s}=-1}$ to $\ket{m_{s}=+1}$ with the help of the state $\ket{m_{s}=0}$.

The fidelity of the population transfer in the nitrogen vacancy center is limited mainly by systematic magnetic errors and dephasing. Systematic magnetic errors and dephasing are the dominant noise for the nitrogen vacancy center and they can be uniformly described by Eq.~(\ref{eq 7}). Systematic magnetic errors in the nitrogen vacancy center result from the imperfect control of the magnetic field used to split states $\ket{m_{s}=\pm1}$. Dephasing in the nitrogen vacancy center is principally caused by the hyperfine interaction with the surrounding $^{13}\text{C}$ nuclear spin bath \cite{noise1,noise2,noise3,noise44,noise31}, which can be described by a random local magnetic field (Overhauser field). Generally, the dynamical fluctuation of the local Overhauser field driven by the pairwise nuclear-spin flip flop is much slower than the typical operate time, making the intensity of the local Overhauser field a random time-independent variable \cite{noise4,noise5,noise3,noise31}. Due to the linear dependence of the states $\ket{m_{s}=\pm1}$ on the magnetic field, the resultant influence of the systematic magnetic errors and dephasing on the nitrogen vacancy center can be seen as frequency errors and therefore can be described by Eq.~(\ref{eq 7}).

As illustrated in Section II, to realize the population transfer from the initial state $\ket{\psi_{i}}=\ket{m_{s}=-1}$ to the final state $\ket{\psi_{f}}=\ket{m_{s}=+1}$ while cancel out the frequency errors to the second-order, we need to find a closed space curve whose tangent vectors at the starting and ending points are along the positive $z$ axis and negative $z$ axis, respectively. Here, we provide a space curve satisfying these conditions and it is constructed as $r(d)=(1-d)r_{1}(d)+dr_{2}(d)$, in which $r_{1}(d)=\sqrt{2}\sin(\pi d)(0,\sin^{2}(\frac{\pi d}{2}),\cos^{2}(\frac{\pi d}{2}))$ , $r_{2}(d)=\sqrt{2}\sin(\pi d)(\cos^{2}(\frac{\pi d}{2}),0,\sin^{2}(\frac{\pi d}{2}))$ and $d\in[0,1]$. The shape of the curve $\textbf{r}(d)$ is shown in Fig.~\ref{Fig3}. It is worth noting that $\textbf{r}(d)$ is not parameterized by arc length, and therefore when using it to calculate the control parameters, one should first transform $\textbf{r}(d)$ into the form parameterized by arc length, i.e., $\textbf{r}(d)\rightarrow\textbf{r}(t)$. The control parameters $\Omega(t)$, $\Delta(t)$ and $\phi(t)$ can be obtained by calculating the curvature and torsion of the space curve $\textbf{r}(t)$. According to Eq. (\ref{eq15}), one can obtain the common Rabi frequency $\Omega(t)$ of the driving fields, which is shown in Fig.~\ref{Fig4} (a). According to Eqs. (\ref{eq16}) and (\ref{eq17}), one can get the information about $\Delta(t)$ and $\phi(t)$. Specifically, if one only changes the detuning $\Delta(t)$ while keeps constant the phase $\phi(t)$ during the evolution, the detuning can be obtained from the torsion of the space curve, which is shown in Fig.~\ref{Fig4} (b). On the other hand, if one only changes the phase $\phi(t)$ of the driving field while keeps constant the detuning $\Delta(t)=0$ during the evolution, the derivative of the phase $\dot{\phi}(t)$ can be obtained from the torsion of the space curve, and correspondingly, by integrating the torsion of the space curve with respect to time $t$, one can obtain the phase $\phi(t)$ of the driving fields, as is shown in Fig.~\ref{Fig4} (c). So, one can choose to change the detuning or phase of the driving fields to realize the population transfer, bringing convenience to the realization in experiment. Without loss of generality, we consider the case of changing the phase $\phi(t)$ while setting the detuning $\Delta(t)=0$ all the time.

\begin{figure*}[tb]
  \includegraphics[scale=0.3]{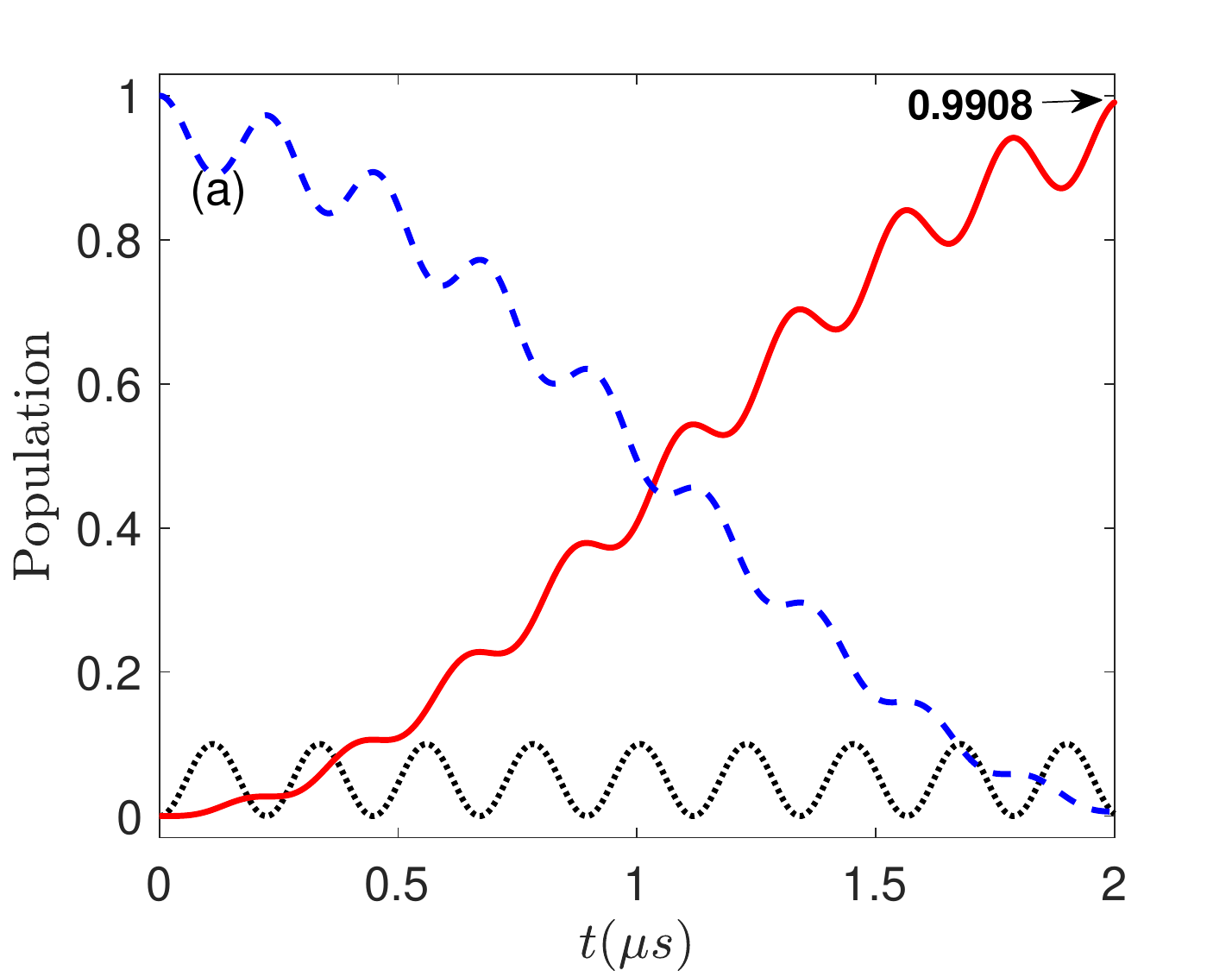}
  \includegraphics[scale=0.3]{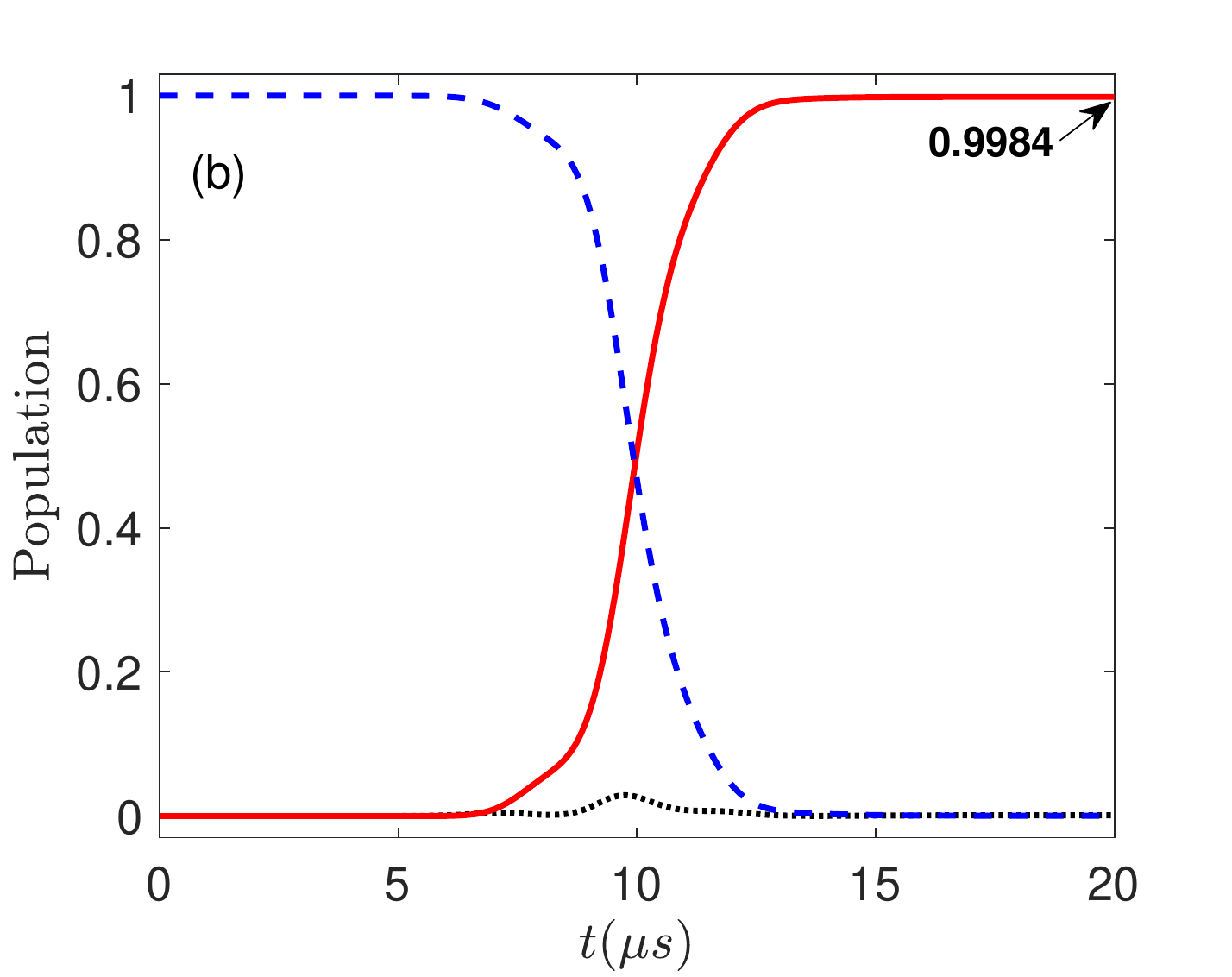}
  \includegraphics[scale=0.3]{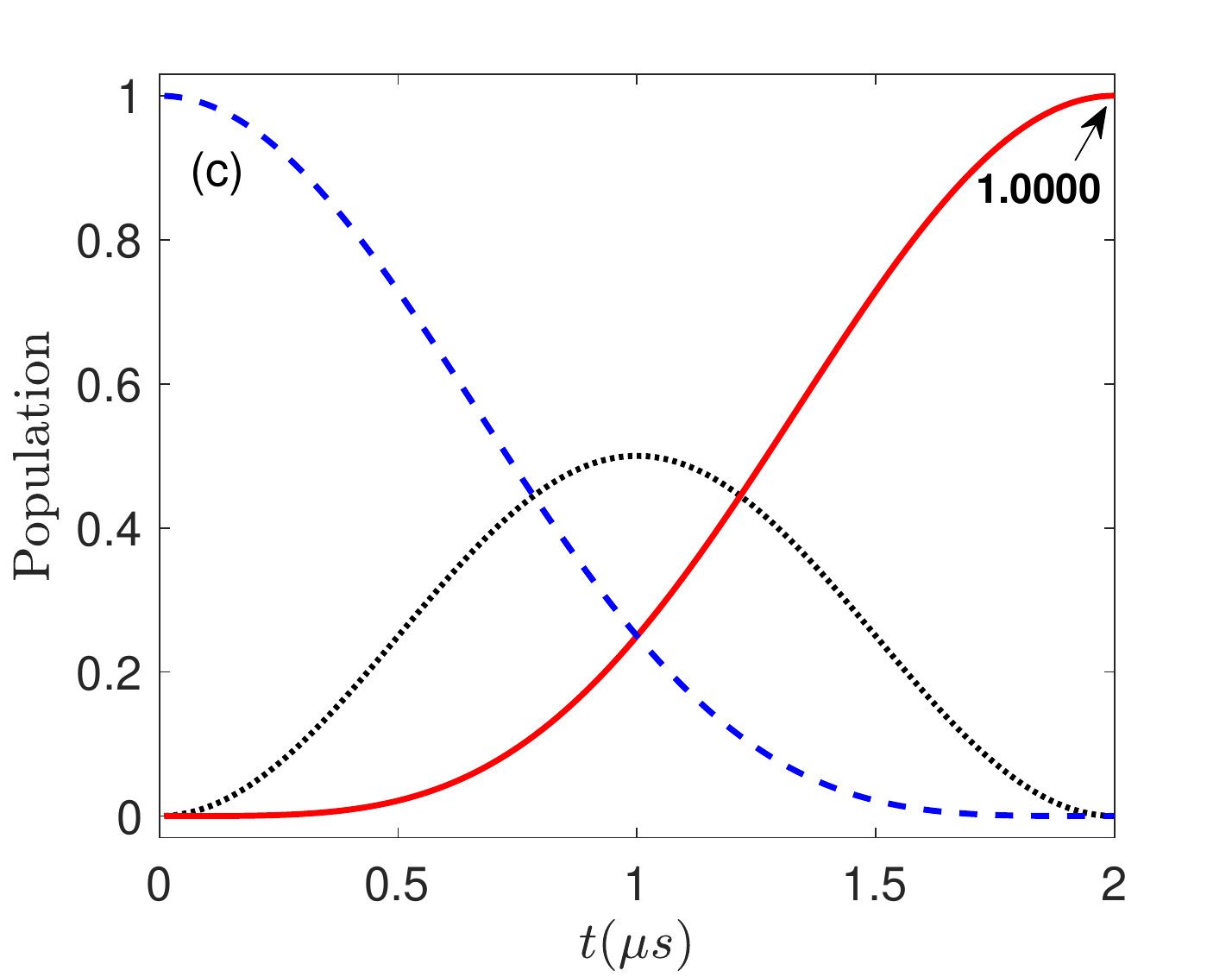}
  \includegraphics[scale=0.3]{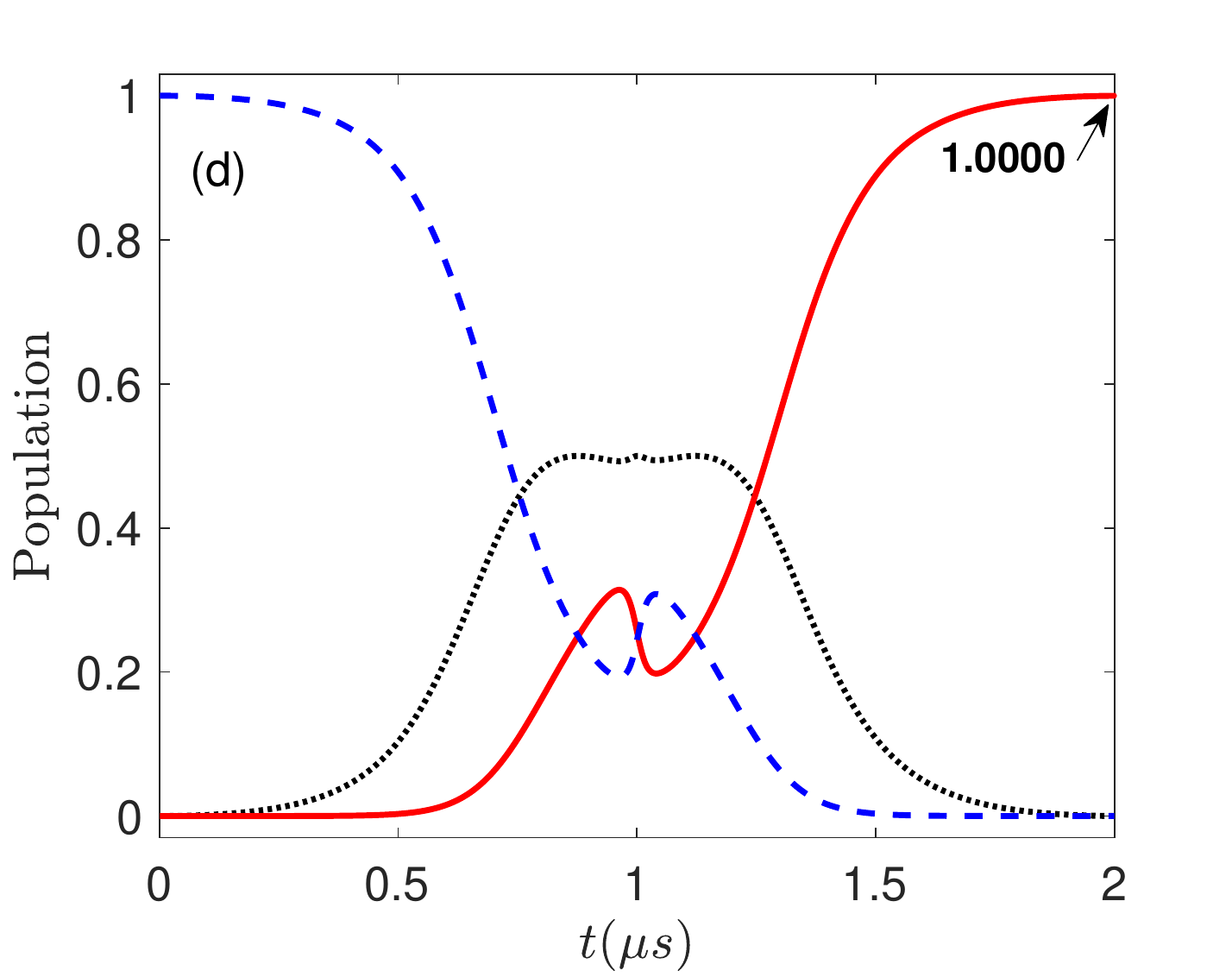}
  \includegraphics[scale=0.3]{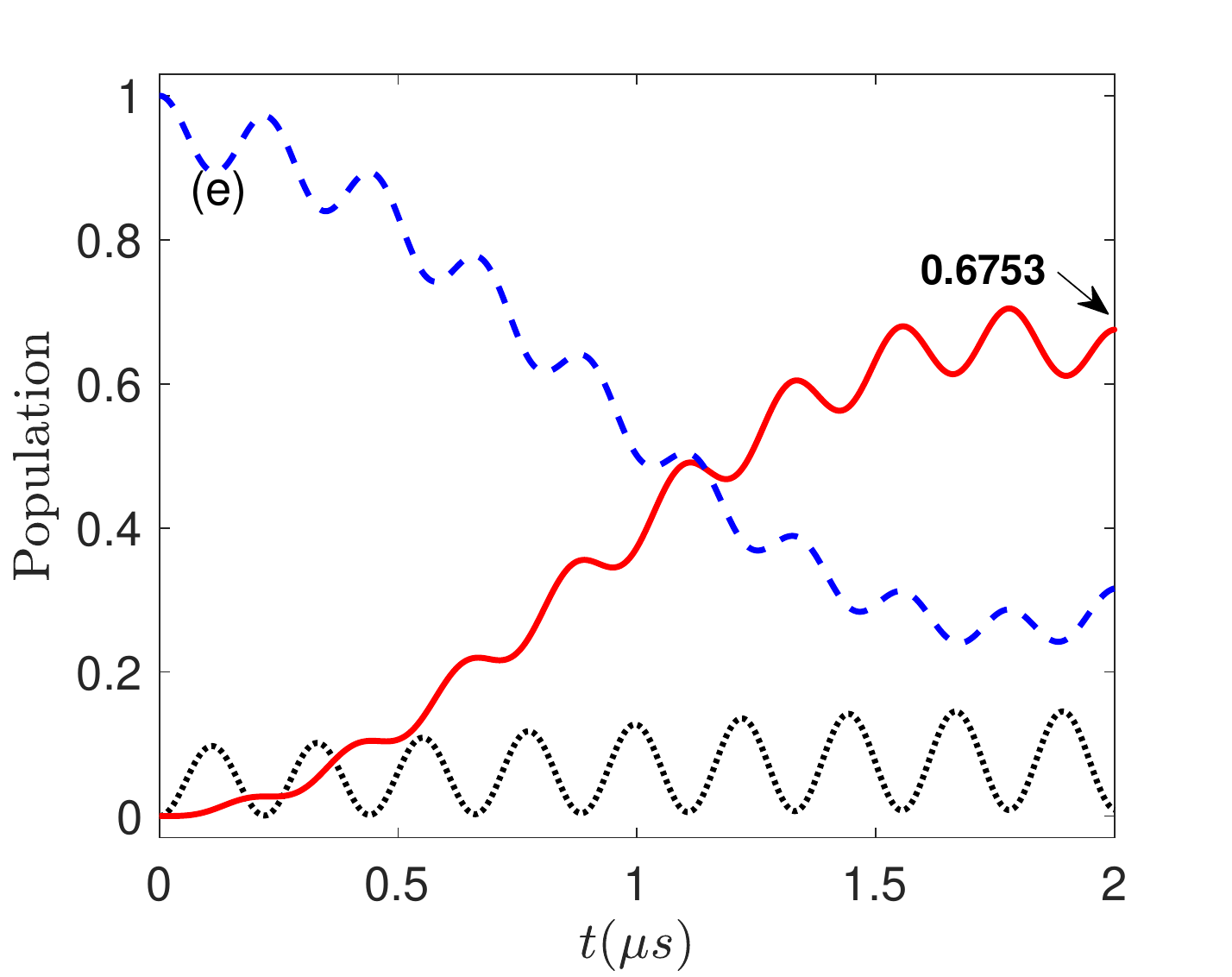}
  \includegraphics[scale=0.3]{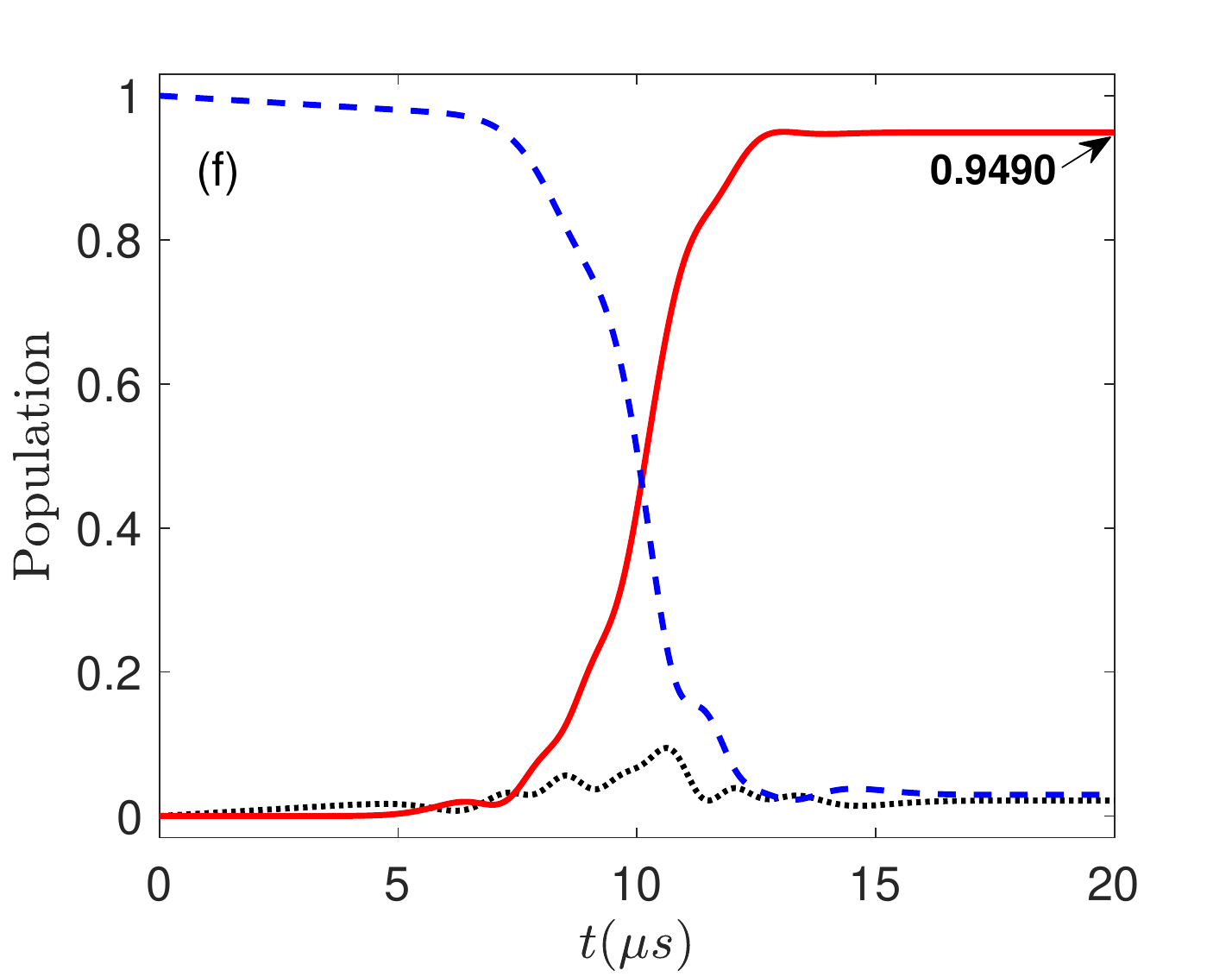}
  \includegraphics[scale=0.3]{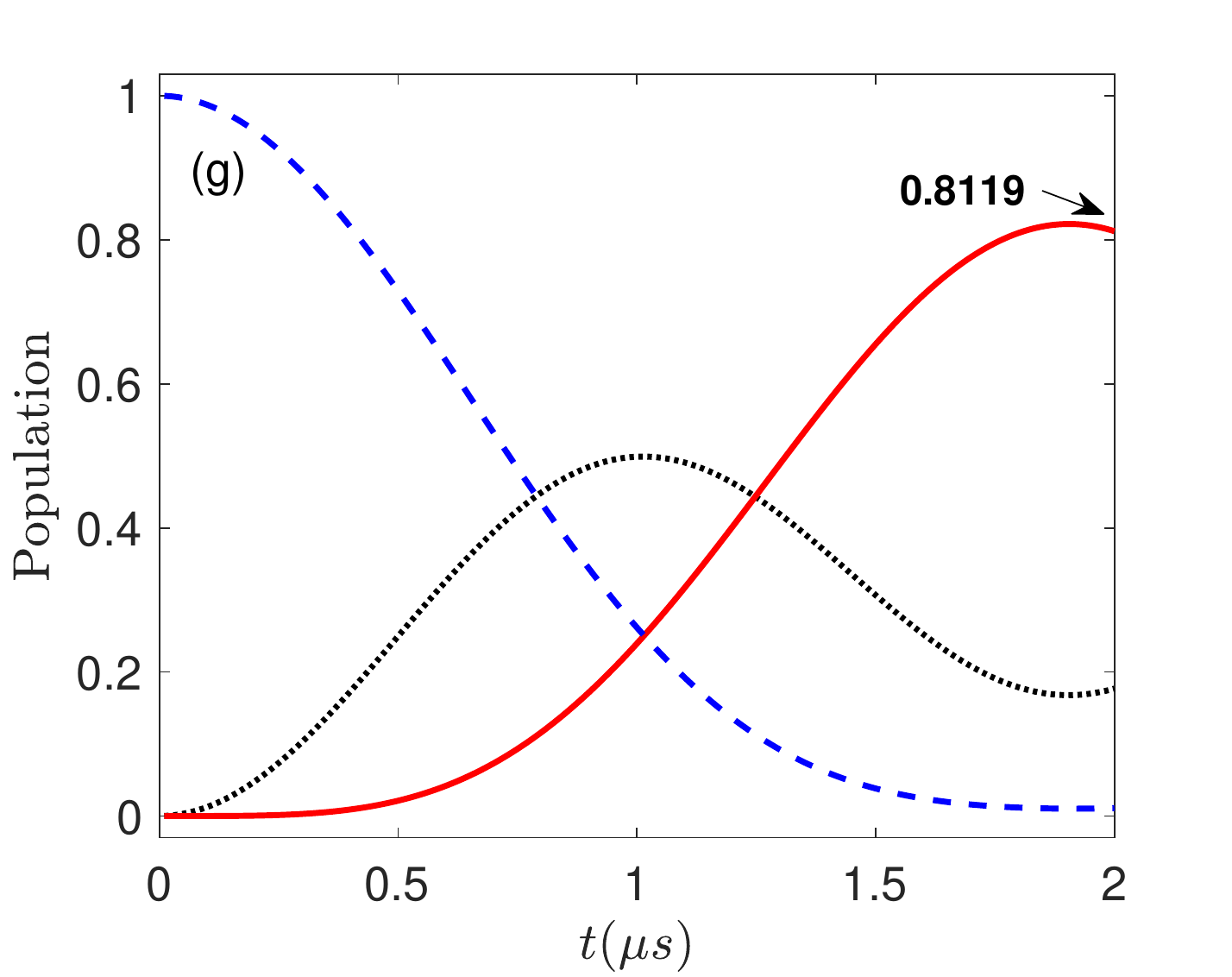}
  \includegraphics[scale=0.3]{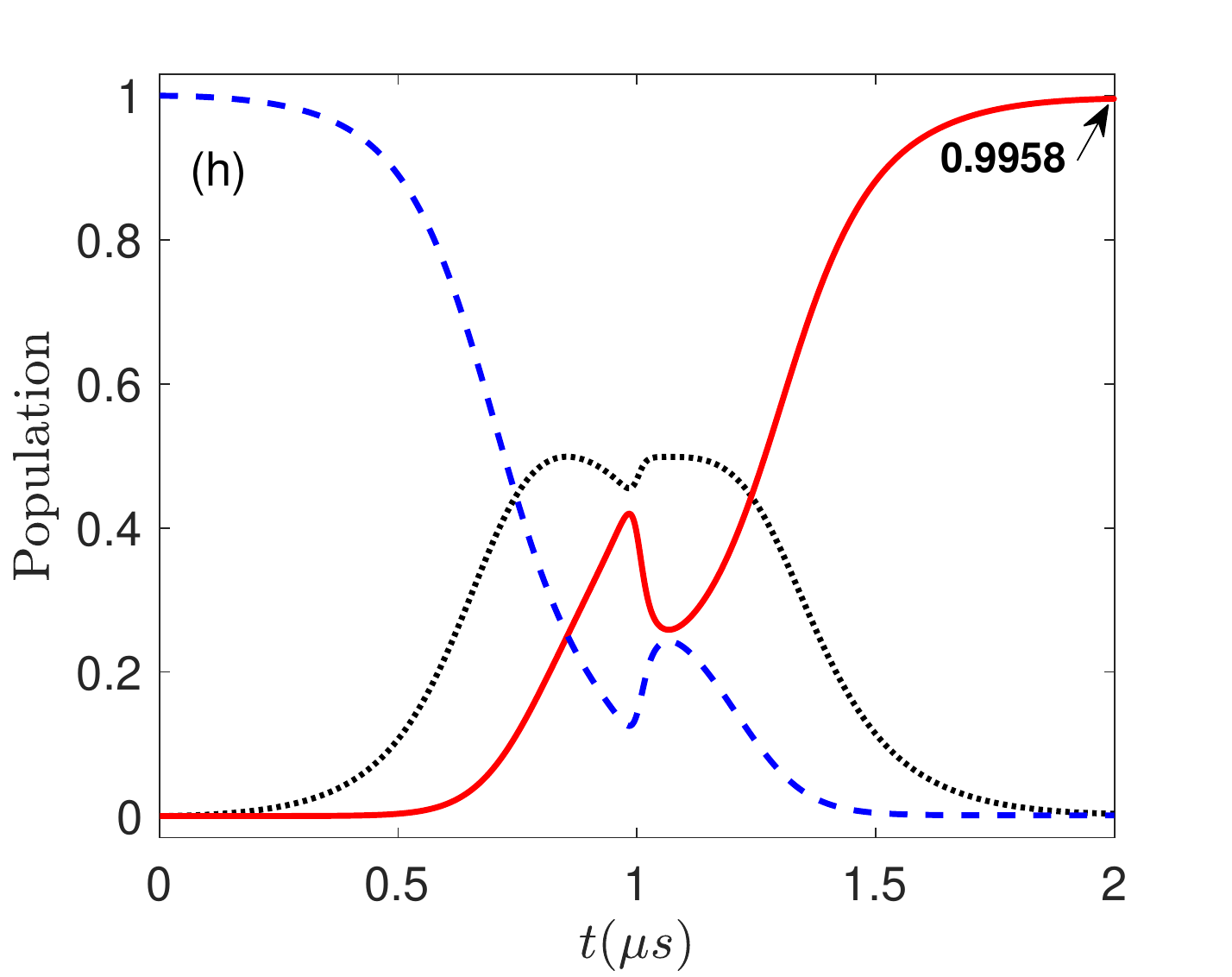}
  \caption{The performance of the population transfers, where the populations of states $\ket{m_{s}=-1}$ (dashed blue), $\ket{m_{s}=0}$ (dotted black), and $\ket{m_{s}=+1}$ (solid red) are presented. The top row shows the simulated results of the ideal case. The bottom row shows the simulated results under the influence of the frequency errors and  the longitudinal spin relaxation process, with $\delta=0.5$MHz and $\Gamma=2$KHz. (a) and (e) The simulation results for SRT with $\Omega_{\textrm{srt+}}=\Omega_{\textrm{srt-}}=2\sqrt{2}\pi$MHz and $\Delta_{\textrm{srt}}=8\pi$MHz. (b) and (f) The simulation results for STIRAP with $\Omega_{\textrm{sti}}=5$MHz, $\Lambda=\mu_{-}-\mu_{+}=3\mu$s and $\sigma=2\mu$s. (c) and (g) The simulation results for the conventional STA based scheme with two resonant driving $\Omega_{\textrm{sta+}}=\Omega_{\textrm{sta-}}=\sqrt{2}\pi/2$MHz. (d) and (h) The simulation results for our scheme, with control parameters $\Omega(t)$ and $\phi(t)$ being obtained from Figs.~\ref{Fig4} (a) and (c), respectively, and proper scaling being implemented to make the total evolution time equal to $2\mu$s.}
   \label{Fig5}
\end{figure*}

To show the efficiency of our scheme, we will numerically simulate the performance of our scheme and compare it with those of SRT, STIRAP and conventional STA based schemes. To make the simulation closer to reality, we consider not only the dominant noise that comes from systematic magnetic errors and dephasing and can be treated as frequency errors, but also the subordinate noise coming from the longitudinal spin relaxation process. To account into both the dominant and subordinate noise, we use the following quantum master equation
${d\rho}/{dt}=-i[H^{\prime}(t),\rho]+\sum_{j,k}\Gamma_{jk}(a_{jk}^{\dag}\rho a_{jk}-\frac{1}{2}\{a_{jk}a_{jk}^{\dag},\rho\})$,
where $H^{\prime}(t)$ is the total Hamiltonian including the frequency errors, and the Lindblad operators $a_{jk}=\ket{m_{s}=j}\bra{m_{s}=k}$ represent the spin relaxation process with rate $\Gamma_{jk}$ corresponding to the longitudinal spin relaxation time $T_{1}$ of the nitrogen vacancy center electron spin. Here we adopt $\Gamma_{10}=\Gamma_{01}=\Gamma_{-10}=\Gamma_{0-1}=\Gamma=2$ KHz, which is proper for nitrogen vacancy centers \cite{cai}.

As mentioned before, SRT, STIRAP and conventional STA based schemes can realize population transfer for uncoupled or weakly coupled spin states. SRT is usually realized by applying two highly detuned driving fields with the intermediate-level detuning $\Delta_{\textrm{srt}}$ and Rabi frequency $\Omega_{\textrm{srt+}}=\Omega_{\textrm{srt-}}$. In the limit of large detunings, $\Delta_{\textrm{srt}}\gg\Omega_{\textrm{srt+}},\Omega_{\textrm{srt-}}$, the intermediate level is scarcely populated, and therefore the system reduces to a two-level system consisting of levels $\ket{m_{s}=+1}$ and $\ket{m_{s}=-1}$ with an effective Rabi frequency $\Omega_{\textrm{srt}}=\Omega_{\textrm{srt+}}\Omega_{\textrm{srt-}}/(2|\Delta_{\textrm{srt}}|)$. Then the population transfer between states $\ket{m_{s}=-1}\leftrightarrow\ket{m_{s}=+1}$ can be implemented approximatively. STIRAP uses two partially overlapping resonant Raman control pulses with the Gaussian envelopes $\Omega_{\textrm{sti}\pm}(t)=\Omega_{\textrm{sti}}e^{-(t-\mu_{\pm})^{2}/2\sigma^{2}}$. The pulse separation $\Lambda=\mu_{-}-\mu_{+}$ and pulse width $\sigma$ is set properly to make the adiabatic condition satisfied. The population transfer between states $\ket{m_{s}=-1}\leftrightarrow\ket{m_{s}=+1}$ can be realized along the adiabatic eigenstate. As a typical example of conventional STA based schemes, we set the parameters in Eq.~(\ref{eq6}) as $\alpha=0, \beta=3\pi/2, \theta(t)=\pi t/2$. By the reverse calculation, we can obtain the Rabi frequencies of the STA control pulses $\Omega_{\textrm{sta+}}=\Omega_{\textrm{sta-}}=\sqrt{2}\pi/2$MHz. With the increase of time, the parameter $\theta(t)$ changes from 0 to $\pi$. At the finally time $t=2\mu$s, the population transfer is realized.

\begin{figure}[htb]
  \includegraphics[scale=0.55]{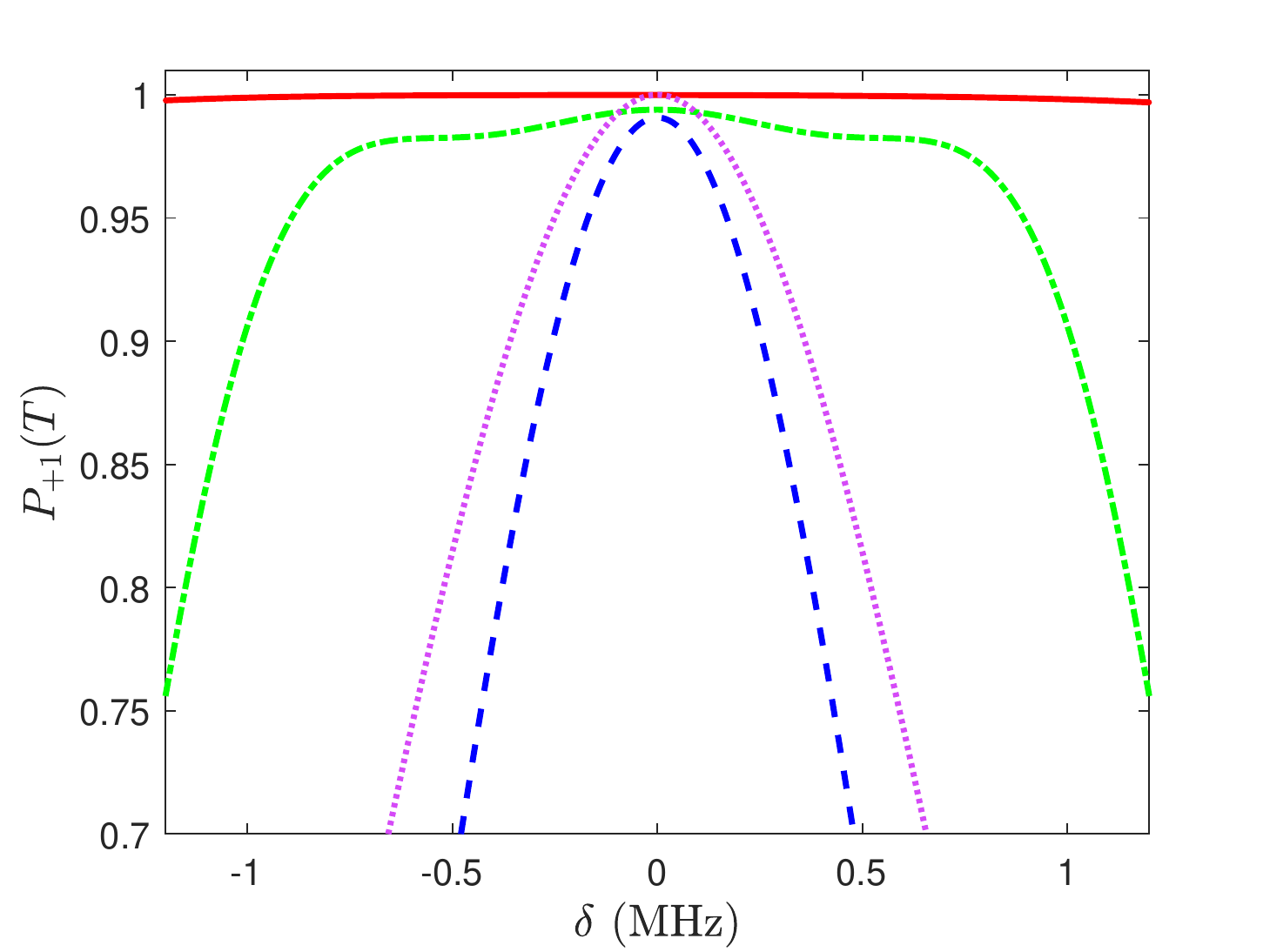}
  \caption{Comparison of the robustness of SRT (dashed blue), STIRAP (dashdotted green), the conventional STA based scheme (dotted purple) and our scheme (solid red) against the frequency errors, with the population $P_{+1}$ of state $\ket{m_{s}=+1}$ at the final time $T$ being the vertical axis, and the strength of the frequency errors being the horizontal axis.}
   \label{Fig6}
\end{figure}

Our simulation results are shown in Figs.~\ref{Fig5} and \ref{Fig6}. Figure \ref{Fig5} shows the performance of the population transfers, while Fig. \ref{Fig6} shows the robustness against the frequency errors. From Figs.~\ref{Fig5} (a), \ref{Fig5} (e) and \ref{Fig6}, one can see that the present of the frequency errors can seriously affect the performance of SRT. From Figs.~\ref{Fig5} (b), \ref{Fig5} (f) and \ref{Fig6}, one can see that STIRAP is partly robust against the frequency errors, the longitudinal spin relaxation process during the longtime adiabatic evolution reduces the fidelity of the population transfer. From Figs.~\ref{Fig5} (c), \ref{Fig5} (g) and \ref{Fig6}, one can see that the conventional STA based scheme is sensitive to the frequency errors. From Figs.~\ref{Fig5} (d), \ref{Fig5} (h) and \ref{Fig6}, one can see that our scheme can achieve high-fidelity population transfer under the influence of both the frequency errors and the longitudinal spin relaxation process. Moreover, Fig.~\ref{Fig6} shows that our scheme is more robust to the frequency errors than SRT, STIRAP and conventional STA based schemes. So, the simulation results show the superiority of our scheme for realizing population transfer under the influence of noise.

\section{CONCLUSION}
In conclusion, we have shown how to realize accurate population transfer between uncoupled or weakly coupled spin states, even under the influence of noise. In our scheme, the population transfer can be implemented fast and it is robust against the frequency errors, the dominant noise in spin systems. Moreover, our scheme is simple to implement. In our scheme, one only needs to find a closed space curve $\mathbf{r}(t)$ starting and ending both at the origin and with the initial and final tangent vectors being $\dot{\mathbf{r}}(0)=(0,0,1)$ and $\dot{\mathbf{r}}(T)=(0,0,-1)$, respectively. The above conditions are not strict so that many space curves would be found. One could choose a well behaved space curve as the candidate and calculate the curvature $\kappa(t)$ and torsion $\tau(t)$ of it to give the control parameters $\Omega(t)$, $\Delta(t)$ and $\phi(t)$. In the above, well behaved means the space curve can give easily realized $\Omega(t)$, $\Delta(t)$ and $\phi(t)$. To show the efficiency of our scheme, we numerically simulate the ground-state population transfer in the $^{15}$N nitrogen vacancy center and compare our scheme with SRT, STIRAP and conventional STA based schemes. The results show that our scheme can still achieve high fidelity under the influence of noise. We hope our scheme can shed light on the accurate population transfer in spin systems.

\begin{acknowledgments}
K.Z.L. and G.F.X. acknowledge the support from the National Natural Science Foundation of China through Grant No. 11775129 and No. 12174224.
\end{acknowledgments}

\end{document}